# Integrating User Experience into Agile

## An Experience Report on Lean UX and Scrum

Manal M. Alhammad
Department of Software Engineering
King Saud University
Riyadh, Saudi Arabia
manalhammad@ksu.edu.sa

Ana M. Moreno
School of Computer Science
Universidad Politécnica de Madrid
Madrid, Spain
ammoreno@fi.upm.es

## ABSTRACT

The integration of Agile development and user experience (UX) has received significant attention over the past decade. The literature contains several process models and wide-ranging discussion about the benefits and challenges of this integration. However, academia has given this integration short shrift. In fact, there are very few publications covering educational courses dealing with Agile development and UX. In this paper, we report on our experience of designing and running a graduate software engineering course that covers the integration of Lean UX into Scrum and employs gamification to improve student engagement. We identified six lessons learned that new point to important aspects to be considered when integrating Agile development and UX in academia. For example, we discuss the complexity of managing Lean UX activities in short sprints or how the design of a testable and tactical hypothesis can be one of the most challenging aspects of the Lean UX process.

## 1 Introduction

While efforts in the integration of Agile and UX practices (Agile UX) goes back a long time [1, 2], the field has received significant attention over the past years in the software industry, pushing forward the frontiers of Agile methods [3, 4]. Research efforts in this field aim to combine the two processes based on the premise that Agile and UX Design (UXD) share common principles, such as iterative development, emphasis on the user, and team coherence. This integration process is still an open issue, because, as discussed later, different challenges and gaps have been identified [3, 5].

From an academic point of view, little is known so far about this integration [6]. On one hand, even though Agile practices are extensively covered by academic courses, very few SE programs include usability [5] and, when they do, core UXD principles and practices are not always properly addressed [6–9]. On the other hand, very few publications have, to the best of our knowledge, specifically reported academic courses that cover the integration of Agile and UX. We believe that the SE education community has a very important role to play in addressing the challenges of Agile UX in industry by educating future software engineers in UXD principles and processes and by highlighting the value of adopting UXD throughout the Agile development process. This is a must in order to introduce this new culture into organizations following a bottom-up approach.

In this paper, we describe a two-year experience in teaching a graduate SE course, designed to introduce students to an integrated approach to developing Agile projects using UX principles and practices. Particularly, we address the integration of Scrum and Lean UX. To do this, we employ a newly developed framework called GLUX (Gamified Lean UX) [10]. GLUX guides the integrated development process using gamification to motivate agile teams to adopt Lean UX practices.

The aim of this paper is to provide useful insights and suggestions for SE educators on how to design and manage the integration of Lean UX with Scrum in an educational context. However, most of the resulting insights may also be applicable in industrial contexts under particular conditions. Generally, beyond contributing to SE education literature by extending existing Agile and UX teaching experiences, we also hope that this paper will contribute to the discussion within the SE community in order to achieve a better understanding of the current trends and challenges of Agile and UX integration.

## 2 Background

### 2.1 Recent Developments in Agile and UX Integration

The integration of UX and Agile software development has been addressed extensively in the literature, leading to the proposal of new methods and techniques for merging the two domains, as well as discussions of the challenges of such integration. Research by Da Silva et al. [3] and Curcio et al. [5] presents a comprehensive and high-level overview of the field, providing an understanding of the evolution of Agile and UX over the past two decades and identifying the different strategies, forms, and challenges of integrating both disciplines. However, most of the integrated processes and techniques published in the literature are rarely followed in practice [3, 10]. The proposed solutions often do not provide a comprehensive integration process that takes into account the organization's culture [12] or team collaboration and communication issues [13]. Nor do they address the lack of basic UX knowledge among software developers, which can be an obstacle to collaboration with UX professionals [11]. Additionally, there are very few empirical studies that evaluate the proposed approaches in industrial and educational contexts. Da Silva et al. [3] stress the need not only to develop guidelines for the integration process, but also to empirically validate those guidelines and approaches. Jurca, Hellmann and Maurer [14] and Curcio et al. [5] confirm that there are few rigorously conducted studies on Agile UX and that most results are inconclusive with respect to the applicability and generalizability of the proposed frameworks in contexts other than the one for which the framework was designed. Wale-Kolade [15] and Bruun and Stage [16] argue that training agile teams in UX methods is essential to address the integration and bridge the gap between software developers and UX professionals.

However, to the best of our knowledge, there are only two studies that discuss the integration of Agile and UX from a training perspective. Péraire [6] presents a Carnegie Mellon University graduate-level course that focuses on integrating interaction design and requirements engineering in the context of dual-track agile, following a mixed approach combining flipped-classroom and traditional delivery. She shares her experience in designing and teaching the course over a four-year period (2015-2019) and discusses the challenges involved in teaching such a course. The other published paper was by Felker, Slamova, and Davis [17], where they report on their experience in integrating UX design with Scrum in the context of a summer research project for undergraduate students. The approach to the integration of UX into Scrum in that study is based on the ad-hoc adoption of a few UX practices, such as contextual inquiry, prototyping, and formative UX evaluation techniques. They also discuss the challenges they faced with this experience and outline a few lessons learned in the form of practical suggestions and guidelines for educators and practitioners when integrating UX into agile. Our paper extends the work of Péraire [6] and Felker, Slamova, and Davis [17] by reporting our experience in designing and teaching a graduate-level course that covers the integration of Agile and UX. Our work differs from [6] in that we address the integration of Agile and UX into a single process, focusing on Lean UX and Scrum, rather than managing parallel interwoven tracks of design and development as in dual-track agile. On the other hand, our work differs from [17] in that we focus specifically on merging Lean UX with Scrum, following a mixed approach of traditional lectures and project-based learning (PBL). We also adopt a novel course design approach by employing gamification as a motivational technique.

### 2.2 Previous Work on Gamification in SE Education and Agile

By definition, gamification employs the philosophy and techniques of game design into non-game contexts to induce a target behavior in people and improve their motivation and engagement [18]. Evidenced by the growing number of publications, gamification in SE is seen as a promising technique to improve software engineers' motivation and engagement and to promote SE best practices [19], [20]. Efforts in this field are motivated by the challenging nature of software development which can entail different practices and processes that are considered repetitive and time consuming, such as code review and bug hunting [19]. While empirical research in this field is still limited, early results paint a positive-leaning picture of the effectiveness of gamification in SE [19, 20]. Most notably, gamification has been implemented in requirements engineering [21], in software project management [22], and in SE education [23]. However, more empirical evidence is necessary to establish reliable conclusions in the SE field [24]. Gamified learning was recognized as a very promising, albeit little explored, approach for improving the learning process and outcomes in SE education [7].

The literature also offers three systematic mappings of gamification applied in SE education [23], Agile [25], and Software Process Improvement [26]. Those studies complement existing research on gamification applied to SE and explore to what extent gamification improves or impacts the motivation and engagement of software engineers. Among the main findings, the authors confirmed that gamification is in its early stages in the field of SE, and very few empirical studies exist. This is mostly consistent with the findings of [19, 20], and [24]. The authors also highlight that gamification is context dependent, and it integrates complex aspects such as human behavior and motivational theories, which makes it far from straightforward. In addition, the authors also raised several critical issues that need to be carefully considered by practitioners who aim to gamify SE processes. For example, gamification should be "integrated into" or "combined with" existing processes or activities; high-intensity gamification, in which the process is overloaded with gamification elements, may have a negative impact, and gamification should not require significant changes that entail a steep learning curve.

## 3 Our Story

### 3.1 Why did we choose Lean UX and Scrum? And why did we gamify the integration process?

*Agile Methods* is a graduate course offered as part of the MS in Software Engineering program at the Universidad Politécnica de Madrid. The aim of this 4-ECTS course is to provide students with an overview of the main techniques used in Agile methods and how they interact with particular quality issues. Before 2019, the course focused on how Agile development processes can account for usability practices to improve the overall quality of the product. This was based on recognizing usability not only as a quality attribute or a non-functional requirement, but also as a crucial factor for delivering successful software products in highly competitive markets [27]. Students were introduced to the main usability heuristics [28], as well as usability guidelines to assure that usability is dealt with from the beginning of the development process [29]. However, students perceived usability as a nice-to-have quality attribute instead of considering it as an essential part of the Agile development philosophy.

In our search for an Agile and UX integration framework for adoption in academia, we discarded the Agile UX approaches published in the scientific literature (journals and conferences), because, as mentioned in Section 2.1, they are not mature enough and, from a practical point of view, are described only briefly in one or two scientific publications where not enough material is provided for an in-depth study and application. We opted instead for Lean UX, which was presented in the book *Lean UX: Designing Great Products with Agile Teams* [30]. Lean UX has a lightweight and iterative nature of the process that is based on design thinking, lean startup, and Agile. As a result, its operation is aligned with Agile development processes.

Additionally, a full and detailed description of the UX practices to be included in the Agile process, as well as specific guidelines and examples on how and when to integrate such practices into a Scrum process are provided in [30]. Consequently, enough suitable material is available. In an attempt to motivate and engage our students to carry out Lean UX activities collaboratively and integrated with Agile, we also employed gamification and designed a framework called GLUX to provide novice Agile teams with effective and engaging help to manage the integration process [10].

### 3.2 Introducing GLUX

GLUX [1] is illustrated in Figure 1 and provides Agile teams with structured and detailed instructions on how Lean UX tactics can be integrated into Scrum [10]. GLUX is primarily designed for small Scrum teams who are struggling with integrating UX activities into the development process, particularly in the absence of UX specialists. Even if a UX specialist is on hand, the GLUX framework can help align the workflow of the whole team towards building user-centered software products and establish a shared understanding of the product throughout the development process. The GLUX framework is divided into two fundamental parts:

a. Five Lean UX tactics to integrate into Scrum: Hypotheses, Design Studio, Experiment Stories, Minimum Viable Product (MVP), and Weekly User Experiments.
b. A customizable gamification strategy based on three game techniques: rewards (points and badges), challenges, and levels.

The key contribution of GLUX is depicted in part *b* of the GLUX framework. The gamification strategy was based on the lessons learned from three systematic mappings of the literature of gamification applied in SE education, agile, and SPI [23], [25], [26]. Our main goal was to explore how gamification was put into practice in the field of software engineering, and to what extent it improves or impacts the motivation and engagement of software engineers. Accordingly, those lessons were taking into consideration throughout the design process of GLUX since they point to several critical aspects that need to be carefully considered when gamifying software engineering processes.

A typical scenario of an Agile team following GLUX is as follows: the team brainstorms and creates a list of hypotheses concerning their assumptions about the needs of potential users, which they discuss with the product owner (PO) during product backlog refinement. The team kicks off the sprint with a design studio in which the team picks a hypothesis as a theme to guide the work of the upcoming sprint/s and start sketching and discussing design ideas accordingly. Immediately after the design studio, the sprint planning meeting should take place. The team decides which user stories they are going to develop and plans for the weekly user experiment in which they put to test a version of their developing product, i.e., an MVP. These plans are captured in what is called an experiment story. The main goal of running weekly user experiments is to validate the team's hypotheses and collect constant feedback to improve the product.

The team should be rewarded for applying each Lean UX tactic and receive additional rewards for doing it collaboratively. The Scrum team is encouraged by special rewards for employing Lean UX tactics for the first time, as well as for rising to some Lean UX challenges.

---

[1] The full documentation of the GLUX framework is summarized at https://bit.ly/GLUXICSE.

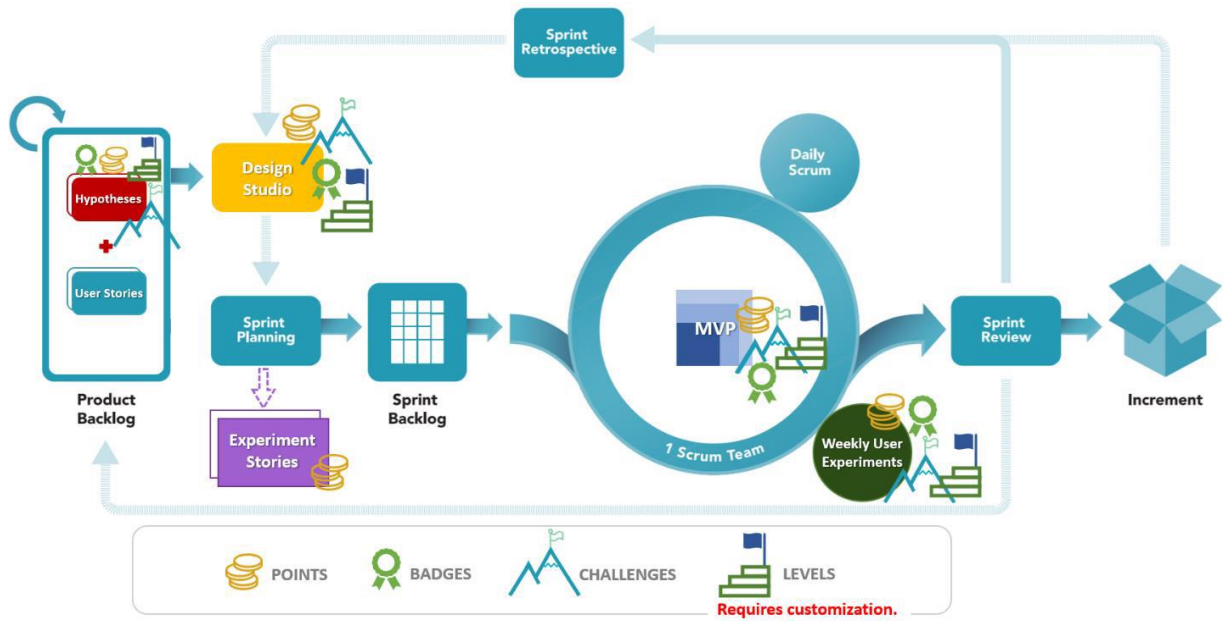

**Figure 1. A general overview of the GLUX framework, integrating Scrum with Lean UX using gamification.**

Challenges are set by the team based on the current difficulties and issues that the team faces regarding UX. The rewards earned should help the team to move up through levels of expertise in Lean UX and are added to the achievements board. The achievements board is a physical or virtual space, where the collected points, badges, and levels of the team are posted. The aim of this board is to visualize the team's achievements and progress in integrating Lean UX tactics.

### 3.3 Course Design

This section describes the course structure and study design of two academic semesters (Fall 2019 and Fall 2020) in which GLUX was implemented.

#### 3.3.1   Learning Objectives

The learning objectives of the Agile Methods course are as follows:
- Understand the agile principles, roles, practices, and artifacts used in Scrum.
- Develop an agile mindset that values early failure, collaboration, continuous learning, continuous improvement, and continuous discovery.
- Establish basic knowledge of Lean UX principles and tactics.
- Learn and practice how UX work can be integrated into the workflow of agile development processes.
- Appreciate the continuous nature of UX and its importance in building high quality software.
Effectively communicate and collaborate with a team.

#### 3.3.2   Overall structure

The Agile Methods course is divided into three parts. The first part is delivered as traditional lectures where the course instructor presents and explains the main Agile development processes and Scrum concepts. The second part focuses on the Lean UX development philosophy and the integration of Lean UX tactics into Scrum. Additionally, the lectures included a few activities like Scrum City using Legos, Planning Poker, and a hands-on workshop to help the students fully understand key Lean UX Canvas concepts [31]. The third part of the course involved developing a team project following GLUX. The course also includes a retrospective class at the end of the semester in which students and instructors conduct an appraisal of the course and the project development process.

We collected data from students through (pre and post experience) questionnaires, unstructured interviews, and the analysis of the students' project reports. The first questionnaire, distributed to the students during the introductory lecture, aimed at performing an initial assessment of the students' Agile and UX knowledge and experience. The second questionnaire was administered during the retrospective class and sought the students' opinions on the process of integrating Scrum and Lean UX. During this last class, we also used unstructured interviews to guide the discussion.

#### 3.3.3   Delivery Format

As mentioned, the course delivery follows a mixed approach, combining traditional lectures with in-class activities, and PBL. In 2019, the lectures and the associated activities were delivered in-class.

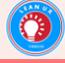

**Figure 2. Three examples of GLUX Cheat Cards, providing a visual summary of GLUX's rules and scoring system.**

However, the entire course was delivered online in 2020, due to the COVID-19 global pandemic. The main challenge for us during online delivery was the management of in-class activities and the course project, particularly the maintenance of student engagement and interaction. Therefore, we adopted a flipped-classroom approach in 2020, where we videorecorded the lectures for the students to watch before the actual lecture time. This left room for interactive group learning activities, such as open discussions and exercises within the virtual classroom.

### 3.3.4 Course Project

In the course project, students simulate a real-world agile project. To do this, they develop an application for managing and funding entrepreneur business ideas, applying Scrum and Lean UX techniques. In the Fall 2019 semester, the 17 students enrolled in the course were divided into three teams. In Fall 2020, the 14 students enrolled in the course were also divided into three teams. Each team was required to develop the project within three one-week sprints. Teams are expected to apply the GLUX framework integrating the Lean UX tactics taught in the lectures (Section 0) into the Scrum process (using the guide discussed in Section 0). Upon adopting these tactics, the teams earn and collect points and badges, and are encouraged to carry out some challenging tasks related to Lean UX. Finally, teams are given the freedom to choose the technological ecosystem to be used in developing their course project and the tools and libraries, for example, to implement user interfaces (e.g. Bootstrap, Vue.js, React.js) or perform user testing (e.g. Maze).

### 3.4 How did our students use GLUX?

As mentioned, the GLUX framework offers a high-level gamification strategy that requires few initializations and customizations, such as assigning numerical values for each Lean UX tactic and designing team-specific challenges, rewards, and levels.

Accordingly, we specifically designed a GLUX guide[2], which illustrates a customized version of the GLUX framework to accommodate our context: three agile teams of graduate students developing a software project using an integrated Scrum and Lean UX process over three one-week sprints. The goal of designing the GLUX guide is to provide a structured process for novice agile teams to integrate Scrum with Lean UX. In addition to the Lean UX tactics, the guide offers additional details on some aspects that might be challenging for novice teams. For example, it explains the different options for a team to manage UX work in the product backlog. In the GLUX guide, we also customized the gamification strategy based mainly on the framework's instructions and included GLUX's cheat cards (Figure 2). Cheat cards are essentially a visual summary of each Lean UX tactic in terms of the rules for integrating such tactics in Scrum, how the scoring system works, and one proposed challenge. Cheat cards made it easier for the students to quickly check the instructions on how to integrate each Lean UX tactic and how the gamified part works.

## 4 FINDINGS AND LESSONS LEARNED

Described in Table 1, this section discusses the six lessons learned we identified from teaching the course for two semesters (Fall-2019 and Fall-2020). They essentially outline the key issues that we came across during the design and delivery of the course (column 1). We also provide a few suggestions on how such issues can be addressed (column 2 and 3). Some of the recommendations are based on our experience, and others were derived from the literature on SE education and Agile UX (in the latter case, the corresponding references are added).

As discussed in section 3.3.2, data collection occurred through two questionnaires (pre and post experience), unstructured interviews, and the analysis of the students' project reports. We followed the qualitative data analysis approach by [32] to synthesize the six lessons learned presented in this section, using NVivo 12 as a supporting tool. The approach by [32] offers a qualitative data analysis process that is based on Grounded Theory (GT). Unlike GT where codes should not be defined a priori and should emerge from the analysis process, the qualitative data analysis approach of [32] starts with "seed categories". Seed categories are an initial set of codes that come from the goals of the study, the research questions, or any predefined area of interest to guides the analysis process. Accordingly, we first formulated a set of "seed codes" that reflect a few aspects we are interested in about the learning process and the overall course material and structure. For example, we were interested to examine any challenge the students had throughout the course, because the course covers intense learning aspects (Lean UX and Scrum), in addition to a course project where students apply those techniques. Thus, "challenge" was included as a seed code.

---
[2] Available at https://bit.ly/GLUXGUIDE2

The next step involved reading all the material and marking where the codes fit to the contents. During this process, some seed codes were reformulated, and others were split into more sub-codes when necessary. For example, the code "challenge" was decomposed into sub-codes to reflect the types of challenges reported about the integration of Lean UX into Scrum such as collaboration challenge, time challenge, and so on. New codes also emerged during the analysis process that represent interesting findings that can help us to understand other aspects. For example, the code "engagement" emerged from reading the students comments about the gamified part of the course to help us understand their perception of the benefits or drawback of including gamification. Formulating new codes and changing existing codes was one part of the analysis where the process was iterative because after that the authors had to go back to re-read and re-code the material again.

After coding the entire dataset, texts of similar and different codes were closely observed and compared accordingly, looking for patterns and themes. For example, we found the challenge of defining a testable and tactical hypothesis was mostly reported during the first sprint, which gave a us the first clue about the steep learning curve issue discussed in section 4.2. The final step of the analysis process involved generating the lessons learned from the conclusions drawn in the previous step.

## 4.1 Integrating UX into Agile can be difficult to manage when working in short sprints

One of the most frequent challenges that we observed during the course and that was constantly reported by our students is the difficulty of fitting the Lean UX process into a short, one-week, sprint. This is a key issue in the case of our students, as there are no separate UX and development teams. One team designs, develops, and validates features in one-week iterations.

The students found that doing development and UX work jointly in one-week iterations was overwhelming, primarily because they are expected to deliver a product increment by the end of the sprint. Lean UX authors envision a typical two-week sprint in their discussions of the integration of Lean UX into Scrum. This gives the team one week to validate ideas through discovery work or implement part of the work without having to deliver a working increment. In our case, however, we were not able to extend the sprint duration due to the course time limits. Almost all teams from both years (2019 and 2020) managed to apply the Lean UX process within the three one-week sprints. However, they had to put in additional effort and allocate extra time to managing the integration process. Some students mentioned that this was an issue, as they found that defining hypotheses and running the design studio took longer than expected. Similar challenges related to not having enough time to carry out UX techniques were also reported in the literature [5].

Josh Seiden, the co-author of the Lean UX book, offers a few guidelines on how to handle UX work accommodated within a sprint [34]. Agile teams need to essentially think of UX work as a continuous part of the process instead of focusing on the finishing line. What cannot be done in one sprint can be broken down into multiple chunks, such that each chunk informs the decisions the team need to make by the end of the sprint. Accordingly, students were reminded in the second round of the course to think of UX as a continuous process, just like development. Additionally, students were instructed to carry out the "just enough" UX work that can inform the team's decisions on a certain feature or functionality. Questions for the team to discuss include: What do we need to validate or learn? How can we do that efficiently with the least waste? What is the most appropriate UX technique to apply here? Nonetheless, students continued to regard the integration of Lean UX into their one-week sprints as difficult to manage.

In terms of the course structure, it might help to redesign the course delivery format to follow a mixed approach combining the flipped classroom with traditional delivery in the form of PBL. In 2020, we had to deliver the course online due to the COVID-19 pandemic, and we followed this flipped-classroom approach. Lectures were recorded and students were asked to watch the videos before class time, so that class time is used for practical exercises. Yet, the short iterations issue was not fully fixed. A longer course duration with a longer sprint length could also help address this issue [7]. The course could be designed to span two semesters, with the second semester fully dedicated for a hands-on development project. This might also give the students the opportunity to engage with real stakeholders as mentors or clients. However, this has not been an option at the Universidad Politécnica de Madrid so far.

Table 1. Lessons learned from our experience in integrating Lean UX into Scrum in an educational context.

| Lesson Learned | Suggested Action – Applied in our Course | Further Recommendations |
|---|---|---|
| **Integrating UX into Agile can be difficult to manage when working in short sprints.** | Remind the students to think of UX as a continuous process [34]. Adopt a flipped-classroom approach so students can gain some practice before the project development. | Design longer courses spanning two semesters to allow for longer sprint length, with the second semester entirely given over to a hands-on development project [7]. |
| **Learning Agile, Lean UX, and the integration of the two processes all at once constitutes a steep learning curve.** | Offer a tailored learning experience based on the current knowledge of the students in agile, UX, and the integration of both processes. | Adopt a fully project-based learning approach to encourage active participation of the students in the learning process [7]. |
| **Defining testable and tactical "hypotheses" is perceived as the most challenging aspect of Lean UX.** | Place more emphasis on the concept of hypothesis (examples and practice). Provide the students with weekly feedback on the team's hypotheses. Adopt a flipped-classroom approach to facilitate interactive group learning activities on the hypothesis concept. | Adopt a systematic process to define and validate hypotheses modeled on hypothesis-driven development [35]. |
| **The high collaboration level of the Lean UX process can be difficult to maintain, particularly among distributed teams.** | Optimize the learning environment for the students to enable real-time collaborative work (e.g., shared workspace, allocated time during the lecture, collaborative tools). | Use gamification mechanisms individually for each team member, instead of group mechanisms. |
| **Integrating Lean UX into Scrum is an abstract process.** | Continuously remind the students that Lean UX and Scrum integration is not a rigid process, and there is always space for flexibility as long as the principles are respected. | Include industry-education engagement activities in the course curriculum (e.g., workshops, involvement of real product owners, hackathons, etc.) to foster students' innovation and autonomy [36]. |
| **When designed properly, gamification can be effective in motivating students to learn and apply the integration of Lean UX into Agile.** | Employ the gamification mechanics, rewards and achievements to engage and motivate the students in the process of integrating Lean UX into Scrum. | Measure the effectiveness of gamification elements and incorporate them gradually [23], [37]. |

## 4.2 Learning Agile, Lean UX, and the integration of the two processes all at once constitutes a steep learning curve

In the first two sprints, students reported that they felt confused and overwhelmed by so many new Agile and Lean UX concepts and practices. The confusion might be logical, as they are dealing with the integration of Agile and UX for the first time with no, or little, prior knowledge and experience (the pre-experience questionnaires confirmed that most students were novice Agile practitioners, whereas they were all new to Lean UX).

Nonetheless, since part of the course was carried out following a project-based approach, we noticed how well the students' knowledge of Agile and UX integration improved sprint after sprint in the course project.

For example, by the third sprint, students defined better hypotheses that were tactical and testable, ran the design studio efficiently, and conducted proper user experiments to validate their hypothesis. This suggests that, although, as expected, integrating Lean UX into Scrum might constitute a steep learning curve with a tough initial learning process, proficiency grows with increased experience. Indeed, we found that the students' knowledge of Lean UX had improved the third sprint.

One thing that might help to make the learning process more effective is to offer a tailored learning experience based on students' current knowledge. We ran a quick survey at the beginning of the course to measure students' knowledge and experience of Agile and UX (and Lean UX in particular). We also made sure that we provided students with constant feedback three times a week, twice via email used by students to provide updates with regard to their daily standups and the current progress of the development process, and then via the sprint review at the end of the sprint. The feedback mostly provided reinforcement of the principles of Agile and UX [13]. Our aim was to offer minimal involvement to encourage students' autonomy and creativity.

The literature on SE education provides few additional strategies that can apply to the context of agile and UX integration. PBL is the main approach used to teach SE trending topics [7]. The active participation of the students in the learning process promotes critical thinking, skills that help the students manage the integration of agile and UX. The benefits of PBL for teaching Agile UX may even be amplified by following a flipped-classroom approach where time is allocated for the students to work on practical exercises collaboratively

### 4.3 Defining testable and tactical hypotheses is perceived as the most challenging aspect of Lean UX

In both courses, we found that all teams had difficulties defining and writing their project hypotheses, particularly in the first two sprints. Students themselves confirmed this issue in the post-experience questionnaires and retrospective class. The hypothesis is one of the core Lean UX concepts. It represents a tactical and testable statement to frame ideas that require validation and experimentation. While there is a template to help teams translate their ideas into hypotheses, students had trouble grasping the concept, particularly the business outcome part.

At first, students thought that this issue might be specific to their limited knowledge and experience with the concept of the hypothesis. This might be true, but the definition of a good hypothesis is not a straightforward activity. This is why the authors of Lean UX designed two canvases as tools to help teams compose and prioritize hypotheses: the Lean UX Canvas [31] and Hypothesis Prioritization Canvas [38], respectively. Both techniques were available for the students to facilitate hypothesis discussion and creation. In fact, a two-hour class was set aside for a hands-on workshop where students worked on composing hypotheses using the Lean UX Canvas. However, we observed that most issues were related to the students' understanding of the hypothesis as a concept and its main components, namely, business outcome, user benefit, and the feature.

We noticed that students having difficulties defining a hypothesis with a business outcome of an appropriate relation or level of detail, like the associated user benefit and feature. In the initial sprints, most teams defined hypotheses that consist of a very high-level business outcome with a very specific functionality, where the functionality itself does not represent a whole feature and may not help the team to achieve the defined business outcome. As we observed during the sprint reviews, the confusion mostly revolved around the difference between a feature and a functionality. Because agile teams are used to thinking of features in terms of user stories that describe specific product functionalities, it was probably challenging to think more abstractly. For example, one team's project idea was similar to the idea of Kickstarter.com where entrepreneurs post business ideas and ask venture capitalists to fund their businesses. This team defined a hypothesis as: *"We believe that we will increase our profit from selling entrepreneurial ideas if investors can invest in ideas using the list ideas feature"*. In this example, the "list ideas" feature is a simple functionality that provides a list of all the ideas posted by the entrepreneurs who need funding. In this case, the business outcome and the user benefit may not be achieved by this particular functionality. Instead, a compound feature such as "invest in ideas" or "fund an idea" may be more suitable. The "list ideas" functionality may form part of a bigger feature along with other functionalities.

The second issue was students defining a hypothesis with a feature that is irrelevant to the business outcome, that is, possibly does not directly contribute to achieving the business outcome. For example, another team's hypothesis states: *"We believe that we can increase the profit margin by 5% if busy users can save time when registering on our website using the one-click signup feature using their Google accounts"*. In this example, "one-click signup" (the feature) may benefit the business in general terms, but it may not directly contribute to "increasing the profit margin" (the business outcome).

The third and final issue was defining a hypothesis with a measurable business outcome. Most teams defined hypotheses with business goals that either cannot be easily measured or were poorly expressed resulting in an unclear real business value. For example, *"we believe that we will increase market awareness of our product if entrepreneurs can promote their business ideas using the share to social media feature"*. In this example, without quantifying or qualifying the level of market awareness by which the specified feature helps to increase it, it would not be possible to determine whether or not the business outcome is achieved.

Allocating more time and resources to explaining the hypothesis tactic may be effective in addressing this challenge. This can be done either during the lectures by providing more specific examples of hypotheses from different contexts or by introducing additional exercises on constructing hypotheses. In the 2020 course, we tried to focus more on explaining the hypothesis tactic through the lectures and provided more examples of different contexts. We also adopted a flipped classroom approach where we provided the students with videos explaining the hypothesis concept and then spent a whole class practicing hypothesis construction and validation using the Lean UX Canvas.

We also gave students weekly feedback on their sprint hypothesis. However, hypothesis creation was again an issue in the second round of the course. We continued to search for additional approaches to provide our students with a more systematic and structured approach to form tactical and testable hypotheses. Hypothesis-driven development [35] offers few tips on how to form testable and tactical hypotheses, such as establishing the metrics to be measured and success factor for benchmarking during hypothesis construction.

## 4.4 A high level of collaboration can be difficult to maintain throughout the Lean UX process, particularly within distributed teams

Our students reported difficulties in collaborating on every UX tactic of the process. The Lean UX process encourages the collaboration of the entire Agile team right from defining hypotheses through to running user experiments. This full-team involvement might be challenging in situations where team members have different working hours/schedules or are distributed across different locations. We observed that most collaboration issues among our students were caused by not being able to arrange face-to-face meetings on a regular basis. The issue of collaboration became even more evident in the 2020 course, when students could not manage to synchronize their meetings and activities due to COVID-19. In 2020, we had 14 students enrolled in the course from different time zones. Students also reported that, although collaboration during the design process was evidently beneficial, it was a challenging activity, as it is better conducted in face-to-face meetings with instant inputs from all team members.

Going back to the principles of Lean UX, it prioritizes collaboration and co-creation to ensure transparency and shared understanding among all team members. The philosophy should remain the same for remote teams or students with different schedules [39]. Educators should aim to optimize the learning environment for the students to enable real-time collaborative work by providing shared workspaces or collaborative tools.

External motivation techniques such as gamification can be helpful in keeping the team engaged in collaborative activities. We rewarded our students with additional points and badges (different from the course marks) for carrying out Lean UX activities collaboratively. This worked well with the class of 2019 when we received positive feedback from the students about the effectiveness of the team-based rewards. However, we received mixed feedback on the same technique in 2020. Students suggested that the points and badges should be given individually to each team member. This way, the active, as opposed to passive, participation of each team member is rewarded.

## 4.5 Integrating Lean UX into Scrum is an abstract process

Our students frequently complained that the Lean UX process is rather abstract, meaning that it lacks specific details on how to manage some activities when integrated into agile. This abstraction caused some frustration and confusion on how to efficiently manage UX work within the sprint. On one hand, all teams from both years stated that they found the process whereby Lean UX tasks are managed in the product backlog confusing. While the Lean UX process recommendations provided some level of detail, they also left room for each team to manage details on task management as per team preferences and objectives. The recommendations suggest either including Lean UX activities as separate product backlog items (PBIs) within the same product backlog together with development work or incorporating each feature's development and UX work combined in one PBI. On the other hand, we noticed in the weekly user experiment tactic that the students did not understand which UX technique should be applied to validate the hypothesis or examine the value of the current feature. Indeed, students sometimes run an experiment whose goal is unclear, or it is not related to the hypothesis to validate.

Moreover, we left it to the students to decide whether to opt for a structured process to run the design studio formally or informally [30]. Some students preferred running the design studio informally because they found themselves communicating and collaborating more efficiently, while others preferred the formal structure as a way to engage the entire team. It may be necessary to continuously remind the students that there are different ways to apply Lean UX within Scrum. We very often had to encourage and help students to improvise a little when they could not figure out how to handle an integration challenge. Then, the students discussed how they managed such challenges and how they improved for the next sprint with the class during the sprint retrospective. Again, it may be useful to provide students with additional resources on UX to foster better decision making on how to run user experiments and collaborate efficiently in managing the design studio. We also think that including industry-education engagement activities can foster the students' innovation and autonomy [36]. In our course, we invited a guest speaker for a special seminar to share her experience in adopting Agile and how her team accounts for usability and UX aspects within the development process. In the future, we plan to host workshops with the collaboration of Agile practitioners and consultants.

## 4.6 When designed properly, gamification can be effective in motivating students to learn Agile UX

Although we received positive feedback on the use of gamification elements to facilitate the integration process, it is important to note that this may not always be the case. Based on observing the students during sprint reviews and analyzing projects reports, we confirmed that designing an impactful gamification strategy is not straightforward. From the feedback received, we found that rewards, particularly points and badges, have the most effective influence on students' engagement and motivation in applying Lean UX in Scrum (see Figure 3). According to the GLUX gamification strategy, badges were earned by the teams, for example, upon carrying out a Lean UX tactic for the first time or achieving a Lean UX challenge. The teams then displayed all the badges on a team achievements board.

This gave the students a more engaging experience integrating Lean UX with Scrum as badges provided visual representation of their team's achievements, while points motivated the students to apply Lean UX tactics more frequently.

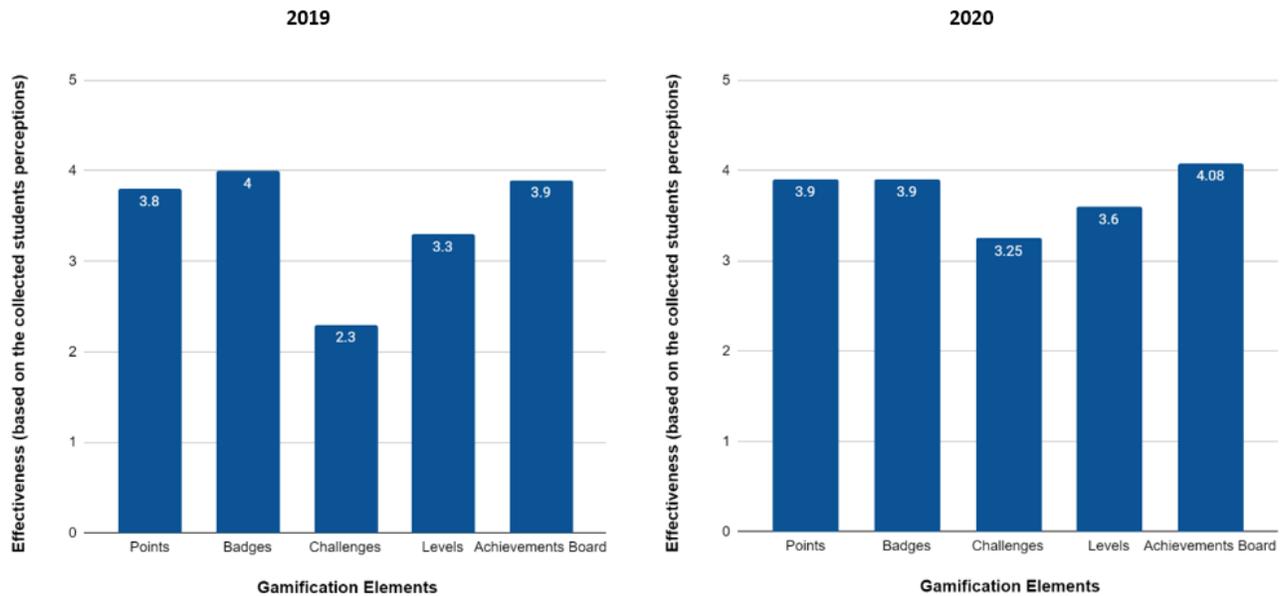

**Figure 3. Students' rating of the effectiveness of each gamification element in improving their engagement and motivation in the Scrum and Lean UX integration process.**

As shown in Figure 3, achievements is another gamification element that was found to be effective at engaging the students in integrating Lean UX with Scrum. As mentioned, GLUX offers a complementary technique called the achievements board, a virtual space where the team posts the points and badges that they earned during the development process. Students enjoyed this technique because it provided them with a dynamic and interactive visual space where they can track and share their awards and provided constant feedback on their progress in applying Lean UX. We also paid particular attention to the extent to which gamification helped the students to improve team collaboration. Our students thought that rewards were helpful in getting the whole team to collaborate.

However, not all the gamification elements were perceived with the same extent of effectiveness in making the process of Scrum and Lean UX integration an engaging process. For example, few students thought that the challenges presented in the GLUX guide added a pressure on the team to apply all Lean UX tactics in all sprints.

Additionally, as mentioned in Section 4.4 above, we noticed a difference in the feedback that we received from the 2019 class and the 2020 class on the reward mechanism. Students of the 2019 class enjoyed the team-based rewards system (where a point or a badge is earned by the team as whole). The students highlighted the benefit of team-based rewards, especially during the design studio tactic, where each and every team member must participate in the design session for the team to earn the associated rewards. On the other hand, students from the 2020 class thought that the rewards should be given individually for them to be recognized and appreciated. The 2020 class think that individual rewards would introduce competition among team members and thus make the process more engaging and fun.

We should note, however, that gamification can be susceptible to the "novelty effect", that is, the excitement and engagement with respect to applying Lean UX tactics may fade away after the first few iterations [37, 40]. This may not represent an issue in university education contexts where a course typically lasts one academic semester. However, it may be necessary to continuously monitor the outcomes of the gamified experience for prolonged experiences.

Either way, gamification should be used as a temporary solution that can help boost the player's motivation towards adopting new habits and learning new skills. The temporary use of gamification is in fact recommended to avoid situations where users end up considering the whole process as a game and focus on collecting rewards and competing against each other, neglecting the primary goal of the original process [40, 41].

Besides, we recommend starting small to avoid wasting time and effort in designing a gamified process that may not work [23, 36]. The gamification elements should be incorporated into the process gradually, starting with one technique, say rewards. Once the applicability of the incorporated gamification element has been validated, additional gamification elements can be used to enforce the same or different objectives.

## 5   Conclusion

The integration of Agile and UX has received significant attention over the past years [3, 5]. Most of the research efforts in this field aimed at examining the different ways of integrating Agile and UX and the challenges of such integration in the software industry.

Meanwhile, the gap between Agile and UX in academia remains significant. In this paper, we reported our experience of designing and applying a novel course that covers the integration of Lean UX into Scrum for graduate SE students. Our aim is to mend the rift between Agile and UX in academia and foster a user-centered development approach. The course follows a hybrid approach of traditional lectures and exercises combined with PBL and employs gamification techniques to improve students' engagement. We briefly explained the structure of the course and how we used a recently developed framework, called GLUX, to guide the students in learning and applying the integration of Lean UX and Scrum. We identified six lessons learned that point to important aspects to consider when integrating Lean UX and Scrum in industrial and educational contexts. We also discuss how gamifying the process of integrating Lean UX into Scrum was well received by our students, as they particularly enjoyed receiving rewards and showing off their achievements.

This paper reports only one experience, and further empirical research is needed to validate the findings and critically evaluate GLUX and, in general, the different methods and techniques proposed in the literature for integrating UX processes and practices into Agile. However, we think that this paper can provide valuable insights for SE academics who are interested in Agile UX courses. Additionally, most of the reported insights are also potentially applicable in the industrial context, as some situations, like software practitioners not having a deep background in UX practices, are also common in this domain.